\documentclass[useAMS,usenatbib]{mn2e}
\usepackage{paralist}
\usepackage{psfrag,color}
\usepackage{graphicx} 
\usepackage{rotating}
\usepackage{txfonts}

\def\nt/f{Nuclear Technology/Fusion}

\title[A distance estimate for the PN NGC~6881]
  {A distance estimate based on angular expansion\\ for the planetary nebula NGC~6881}
\author[L. Guzm\'an-Ram\'\i rez et al.]
  {Lizette Guzm\'an-Ram\'\i rez,$^1$\thanks{lizette.ramirez@postgrad.manchester.ac.uk}
    Yolanda G\'omez,$^2$
 Laurent Loinard,$^2$ 
Daniel Tafoya$^3$\\
   $^1$Jodrell Bank Centre for Astrophysics, University of Manchester, Manchester M13 9PL, UK\\
  $^2$Centro de Radioastronom\'ia y Astrof\'isica, Universidad Nacional Aut\'onoma 
de M\'exico, 58089 Morelia, Michoac\'an, M\'exico \\
  $^3$Department of Physics and Astronomy, Graduate School of Science and Engineering,
Kagoshima University, 1-21-35 Korimoto, Kagoshima 890-0065, Japan
}
\date{Released 2011 Xxxxx XX}

\pagerange{\pageref{firstpage}--\pageref{lastpage}} \pubyear{2011}

\def\LaTeX{L\kern-.36em\raise.3ex\hbox{a}\kern-.15em
    T\kern-.1667em\lower.7ex\hbox{E}\kern-.125emX}

\begin{document}

\label{firstpage}

\maketitle
 
\begin{abstract}
In this paper, we report on high angular resolution radio observations 
of the planetary nebula NGC~6881 obtained with the Very Large Array 
at a wavelength of 6~cm. The emission appears to be the superposition 
of a roundish core and a point-symmetric bipolar structure elongated 
along a position angle of about 145$^\circ$. This is strongly reminiscent
of the morphology seen in H$\alpha$ and [NII] images. A comparison
between VLA observations obtained in 1984 and 1994 clearly reveals the
expansion of the core of the nebula, at a rate of 2.1 $\pm$ 0.7 mas~yr$^{-1}$. 
Assuming that the expansion velocity in the plane of the sky (determined 
from these measurements) and the expansion velocity along the line of 
sight (estimated from optical spectroscopy available in the literature) are 
equal, we find a distance to NGC~6881 of 1.6 $\pm$ 0.5 kpc $\pm$ 0.3 kpc, where
the first error reflects the uncertainty on the expansion, and
the second error comes from the potential difference between
pattern and material speeds. This distance
is compatible with (but does not necessarily imply) an association of 
NGC~6881 with the nearby HII region Sh~2-109 and, more generally, 
the Cygnus star-forming region. 
\end{abstract}


\section{Introduction}
NGC~6881 (PG~74.5+02.1, IRAS~20090+3715) is a very interesting planetary nebula 
(PN) with a system of multiple bipolar lobes traced by the emission from ionized as well 
as neutral material \citep{gm98,gue00,ks05,rlgm08}. Originally, \cite{gm98} classified it 
as a quadrupolar planetary nebula based on the examination of the [N II] emission and 
the [N II]/[O III] and [N II]/H$\alpha$ ratio images. These authors found one pair of lobes 
consisting of two highly collimated structures that finish as bright knots where the emission 
of [NII] is enhanced (outer lobes). With a size of about  8$''$, the inner lobes are shorter 
but much brighter than their outer counterparts. Moreover, they exhibit a clear point-symmetric 
morphology. The expansion velocity derived for these lobes is  90~km s$^{-1}$, which 
yields a kinematical age of about 400$d$ years (where $d$ is the source distance in 
kpc). 

\cite{ks05} obtained Hubble Space Telescope images of this nebula from which they 
identified one more pair of ionized lobes. They have the same extension and morphology 
as the inner lobes but with a position angle slightly different. This discrepancy is not well 
understood. These authors also found several rings circumscribed in the inner walls of the 
ionized bipolar lobes. The presence of these multipolar lobes and rings oriented in different 
directions suggests the action of precessing collimated fast outflows that change their 
precession axis as a function of time. \cite{gm98} proposed this to explain the existence 
of a loop-like structure traced by the ionized emission, located at the southeast edge of the
lobes and elongated in the direction perpendicular to the lobes. They modeled the loop as 
a precessing jet that, assuming ballistic expansion, started $\sim$7400$d$ years ago. 
However, \cite{rlgm08} found that the location of the loop-like structure coincides perfectly 
with the ending tip of the southeast H$_{2}$ lobe, suggesting that it is marking a region of 
interaction of the bipolar lobe with dense material. These authors noted that NGC~6881 
is projected towards the HII region Sh~2-109 and suggested that they are physically 
associated. They proposed that the southeast lobe of NGC~6881 is interacting with dense 
material, forming the observed loop-like structure and the sharp edge of the H$_{2}$ lobe. 
The HII region Sh 2-109 is located at a distance of 1.4$\pm$0.4 kpc \citep{fic84}.  Using a 
distance scale proposed by \cite{cah92}, \cite{rlgm08} calculated a distance of $\sim$1.5 
kpc to NGC~6881, which supports the association with the region Sh~2-109. However, the 
scale developed by \cite{z95} yields a distance of 6.4 kpc, in which case they would not be 
physically associated.

Statistical techniques can provide distance estimates to PNe as accurate as 50\% on average, 
but when they are applied to individual objects the errors can be as large as a factor of 2 or 
even more \citep[e.g.][]{taf11}. This results in a very uncertain estimation of  the Galactic location 
and physical parameters of PNe. In order to obtain more accurate estimations of the distance to 
PNe, other techniques have been used. One that has proven effective is the so-called expansion 
parallax method, which is based on the comparison of  the angular expansion of the source on 
the plane of the sky, with the velocity expansion along the line of sight measured from the width 
of some appropriately chosen spectral lines \citep{mass86}. In principle, the angular expansion 
could be measured using any tracer, but the free-free radio emission from the ionized gas has 
been, by far, the most popular choice \citep[][]{mass86,mass89,gom93,htb93,htb95,km96,hat96,
cs98,ggr06,zhp08,gglm09}. Hubble Space Telescope observations at optical wavelengths have 
also occasionally been used to measure the expansion \citep[][]{rbhk99,pbht02}. 

In this work we have used the expansion parallax method to measure the distance to the PN 
NGC~6881. Our new distance estimate will be used to determine with improved accuracy the
physical parameters of the source, particularly its size and age. It will also enable us to examine
the association of NGC~6881 with the HII region Sh 2-109. For this purpose, we have retrieved 
data from the Very Large Array (VLA) archive of the National Radio Astronomy 
Observatory\footnote{The NRAO is operated by Associated Universities, Inc.\ under a cooperative 
agreement with the National Science Foundation.}  (\S 2). The structure of NGC~6881 and the 
details of the component that was used for the measurements are described in \S 3. The technique 
used to determine the angular expansion is presented in \S 4. The estimation of the distance and 
the location of NGC~6881 in our Galaxy, as well as other physical parameters, are discussed in 
\S 4.1.  

\section{The data}

We have made use of two data sets of observations toward the PN NGC~6881 retrieved from the 
VLA archive.  The observations were performed in the continuum mode (bandwidth $=$ 50 MHz) 
at a frequency of 4.8~GHz (corresponding to a wavelength of 6~cm). The configuration of the array 
for both data sets was the most extended one (configuration A). The two data sets correspond to 
observations carried out on 1984, December 29 (1984.99; project AK113) and 1994, March 31 
(1994.24; project AH509), respectively. This gives a time baseline of 9.25 years between the 
observations. The data were edited and calibrated using the Astronomical Image Processing 
System (AIPS) following standard procedures (see Table 1). Both data sets were self-calibrated 
in phase and amplitude.

\begin{table}
\caption{Observational parameters}
\begin{tabular}{ccc}
\hline
Epoch    &  Phase Calibrator &Bootstrapped Flux density \\
\hline
1984 Dec 29 (1984.99)&2023+336&2.07$\pm$0.01 Jy \\
1994 Mar 31 (1994.24) & 2005+403 & 2.98$\pm$0.01 Jy \\
\hline
\end{tabular}
\end{table}

\section{The structure of the PN NGC~6881}

\begin{figure}
\centering
\includegraphics[angle=0,width=\columnwidth]{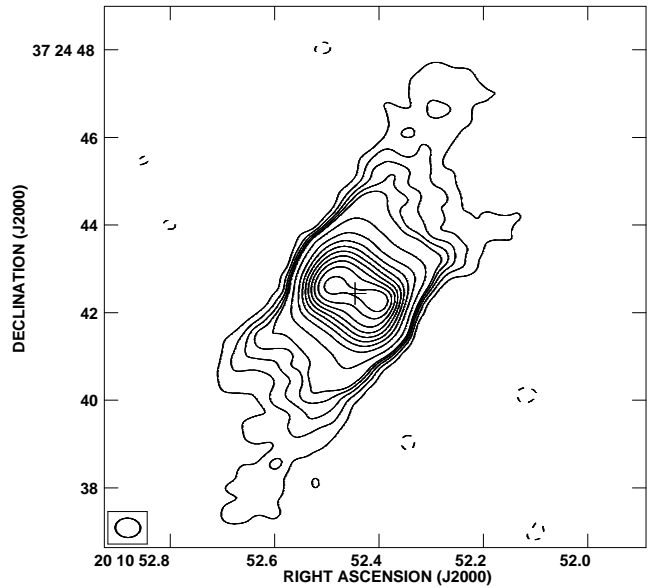}
\caption{Contour image of the 4.8 GHz (6~cm) continuum emission from 
NGC~6881 for the 1994.24 epoch. The contours are  -1, 1, 2, 3, 4, 5, 10, 
20, 30, 40, 50, 60, 70, 80, and 90$\%$ the peak flux of 9.8 mJy~beam$^{-1}$. 
The $rms$ noise of the image is 0.025 mJy~beam$^{-1}$. The synthesized 
beam  ($0\rlap.{''}57 \times 0\rlap.{''}44$ with a position angle of 87$^\circ$) 
is shown in the bottom left corner of the image.The cross marks the position 
adopted as the center of the nebular emission: $\alpha(2000)$= 20$^h$ 
10$^m$ 52$\rlap.{^s}$446, $\delta(2000)$= $+$37$^\circ$ 24$'$ 
42$\rlap.{''}$43.}
\label{fig1}
\end{figure}

\begin{figure*}
\centering
\includegraphics[scale=0.70,angle=0]{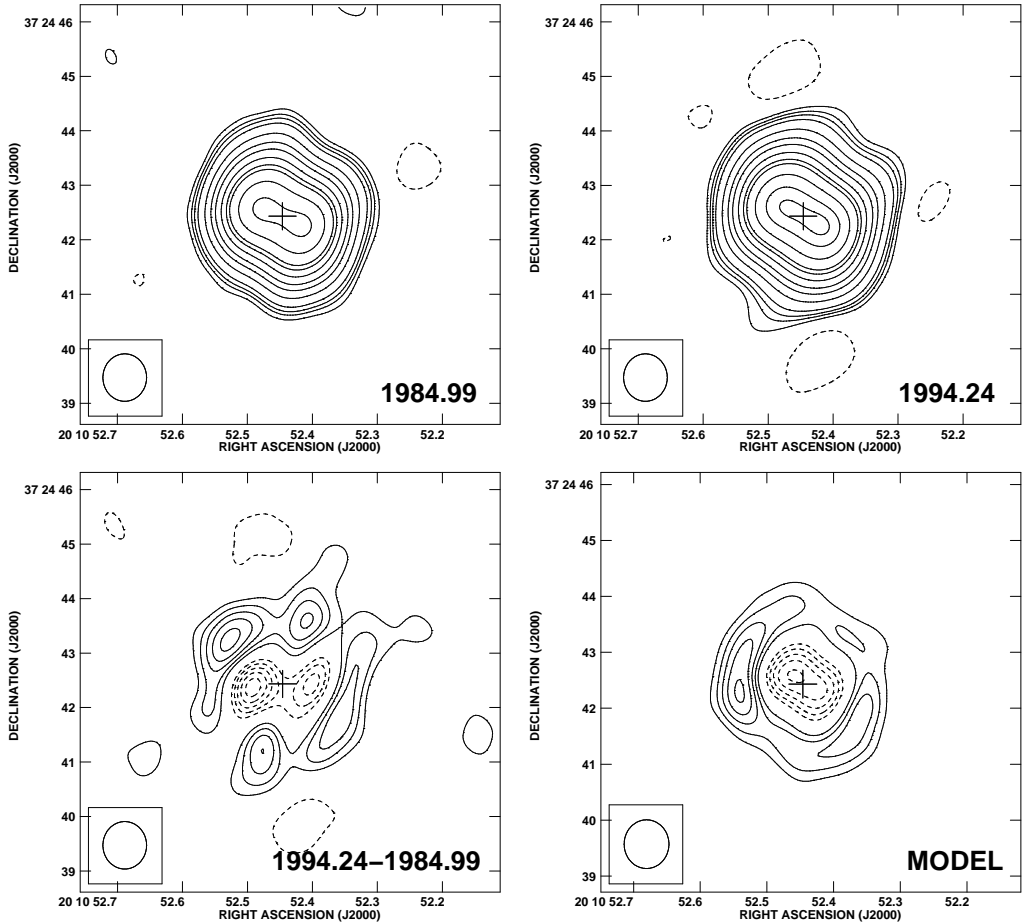}
\caption{Top: Reconstructed contour images of the 6 cm continuum
emission from NGC 6881 for 1984.99 (left) and 1994.24 (right).
The contours are  $-$4, 4, 7, 10, 15, 30, 50, 70, 100, 
130, 150, 200, 250, and 280 times 0.07 mJy~beam$^{-1}$, the average 
$rms$ noise of the images.The individual $rms$ of the images is 0.09 mJy~beam$^{-1}$ for the 1984.99 image and 0.05 mJy~beam$^{-1}$ for the
1994.24 image. 
Bottom: contour images of the 6~cm difference image
(left) and of the best model (right) obtained as described in the text.
The contours are $-$11, $-$9, $-$7, $-$5, $-$3, 3, 5, 7, 9, and 11 
times 0.095 mJy~beam$^{-1}$, the $rms$ noise of the difference image.
The restoring beam ($0\rlap.{''}87 \times 0\rlap.{''}80$
with a position angle of $0^\circ$) is shown in the bottom left corner
of each image. 
}
\label{fig2}
\end{figure*}

In the radio continuum images, NGC~6881 clearly exhibits a bipolar, point-symmetric structure 
elongated along a position angle of 145$^\circ$ (see Fig. 1). This morphology is very similar 
to that of the H$\alpha$, [NII] and [OIII] images presented by \cite{gm98}, although the radio 
continuum emission image is not as sensitive in the outermost portions of the lobes as those 
in H$\alpha$ and [NII]. In consequence, neither the knots nor the loop-like structure, which are 
detected in [NII] light, are detected in radio continuum emission. In the central region of the 
nebula, a two-peaked bright core (also seen in [NII] images) elongated along a position angle 
of  55$^\circ$ is clearly present. Guerrero \& Manchado (1998) interpreted this structure as an 
expanding ring in the equatorial plane of the nebula. 

If we assume that the expansion velocity of the central core along the line of sight is the same 
as that on the plane of the sky, the angular expansion rate should be about 2 mas~yr$^{-1}$
if the source is at 1.5 kpc and about 0.5 mas~yr$^{-1}$ for $d$ = 6.4 kpc. Such values, although 
small, are measurable, provided the core of the emission can be separated from the lobes. For 
interferometric data, the structures associated with different angular scales can be separated 
by restricting the portion of the $(u-v)$ plane considered during the imaging process (G\'omez 
et al.\ 1993). In the case of NGC~6881 the emission from the extended lobes can be almost 
entirely suppressed by selecting only the visibilities corresponding to $(u-v)$ spacings longer 
than about  30 k$\lambda$ (corresponding to $\sim$7${''}$). In addition, the expansion is best 
detected if the longest $(u-v)$ spacings are somewhat down-weighted and the angular 
resolution is degraded so the source is only a few resolution elements across (e.g.\ G\'omez 
et al.\ 1993). In the present case, this was achieved by considering only visibilities in the $(u-v)$ 
range from 30 to 550 k$\lambda$, and by applying a Gaussian taper function with a full width 
at half maximum of 200 k$\lambda$. Finally, the images were reconstructed to minimize possible differences in the u-v coverage using a beam of $0\rlap.{''}87 \times 0\rlap.{''}80$, P.A.= $0^\circ$, the average of the individual beams of each 
observation. These procedures yielded noise levels of $\sim$ 0.09 and 0.05 mJy~beam$^{-1}$, 
for the 1984 and 1994 data sets, respectively. The difference image between the two epochs 
(1994.24$-$1984.99), was produced following Guzm\'an et al. (2006, 2009), and is shown in 
the bottom left panel of Figure~2. This image shows the typical negative and positive pattern 
expected when expansion is present.

\section{Estimation of the angular expansion}

The images of NGC~6881 obtained at epochs 1984.99 and 1994.24 (upper panels of Figure 2) 
look very similar, so the expansion of the nebula is not immediately obvious. The negative-positive 
pattern seen in the difference image (1994.24$-$1984.99), however,  clearly reveals the expansion 
of the ionized nebula.   \citet{gom93}, and more recently \cite{ggr06,gglm09} have developed a
technique that has proven effective to measure very small expansion rates using observations
at two epochs. The technique is based on the assumption that the expansion of the ionized nebula 
is self-similar. To characterize the expansion, the image corresponding to the better of the two data 
sets is self-similarly expanded by a factor $1+\epsilon$ (if the best data set corresponds to the older 
observation) or shrunk by a factor $1-\epsilon$ (if it is the more recent), and subtracted from itself. 
This provides a model difference image that can be compared with the true difference image. The
correct expansion rate clearly corresponds to the value of $\epsilon$ for which the model difference 
image is most similar to the measured difference image. Quantitatively, this corresponds to the
value of $\epsilon$ that minimizes the $\chi^2$ defined as

\begin{equation}\centering
\chi^2 = \sum_{i,j}(M_{i,j} - T_{i,j})^2/\sigma^2,
\end{equation}

\noindent where $M_{i,j}$ is the value at pixel $(i,j)$ in the model
difference image, $T_{i,j}$ is the value at the corresponding  pixel
of the true difference image and $\sigma$ is the standard deviation of the 
data. The sum is taken over the appropriate portion of the images.

In the present case, the data obtained at epoch 1994.24 have better quality, so the 
corresponding image was contracted by a factor (1 $-$ $\epsilon$). The value
of the $\chi^2$ as a function of $\epsilon$ is shown in Figure~3. It has a clear
minimum at $\epsilon$ = 0.027. The model difference image corresponding to that 
value of $\epsilon$ is shown in the bottom right panel of Figure 2. Once the minimum 
value of $\epsilon$ is determined, the confidence limits on this value are estimated as
follows. The number of independent data points in the fit, $n$, is estimated by dividing 
the solid angle of the region used to estimate $\epsilon$ by the solid angle of the 
synthesized beam. Here, the region of the images considered to calculate the $\chi^2$
corresponds to $n = 20$ beams. Assuming that the plot in Figure 3 represents a normal 
$\chi^2$ distribution and following \citet{press92}, we scale 
by multiplication the value of the minimum to be $20$. The confidence 
limits 
are estimated by determining the $\epsilon$ interval given for a value in the horizontal 
axis of $n+1$ = 21 \citep{avni,wall96}. This 
yields $\epsilon = 0.027 \pm 0.010$.

\begin{figure}
\centering
\includegraphics[scale=0.4,angle=0]{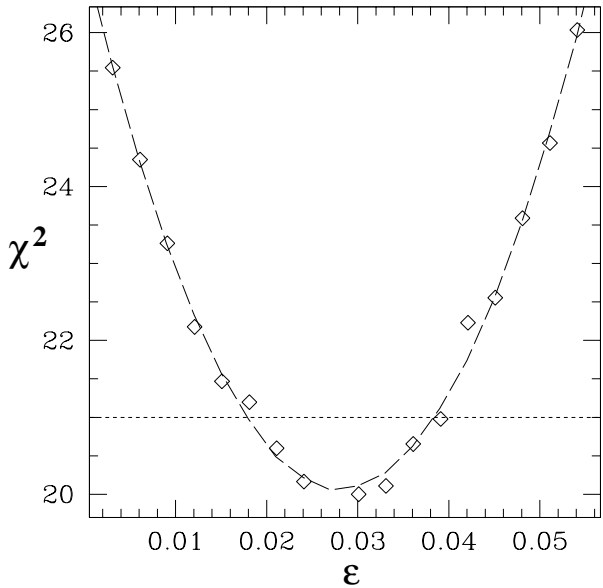}
\caption{$\chi^2$ fit of the residual image obtained from 
subtracting the model to the data as function of the 
contraction factor $\epsilon$. The minimum of the fitted curve (dashed-line) 
indicates the value of $\epsilon$ that minimizes the 
$\chi^2$ function.}
\label{fig3}
\end{figure}

\begin{table}
\caption{Physical parameters for NGC 6881}
\begin{tabular}{lr}
\hline
Parameter    &  Value \\
\hline
Flux (4.8 GHz) [mJy]& 120.0 $\pm$ 0.4 \\
Distance [kpc] & 1.6 $\pm$ 0.5 \\
Radius of the central ring [AU] & 1170 $\pm$ 380 \\
Expansion rate of the ring [km~s$^{-1}$] & 16 $\pm$ 2 \\
Kinematic age of the ring [years] & 350 $\pm$ 110 \\
$n_e$ in the central region  [$10^4\;{\rm cm}^{-3}$] & 5 $\pm$ 1  \\
$M_i$ in the central region  [$10^{-2}\;M_{\odot}$] & 0.22  $\pm$ 0.08 \\
EM in the central region [$10^6\;{\rm cm}^{-6}{\rm pc}$] & 28  $\pm$ 1\\
\hline
\end{tabular}
\end{table}

The good agreement between the images produced for the real difference and the 
model difference suggests that the assumption of self-similar expansion for the central 
ring of the PN NGC~6881 is reasonable, at least in the equatorial plane of the nebula. The structure in 
the direction of the bipolar lobes appears to be more 
complex, and a different expansion velocity ($\sim$90 km~s$^{-1}$; Guerrero \& 
Manchado 1989) might be more appropriate. Moreover, as we mentioned earlier, the 
outflow direction may be changing with time (Kwok \& Su 2005). Therefore, a 2D or 
3D model (such as that presented by Zijlstra et al. (2008) for NGC~7027) would, in 
principle, be needed. However, constructing a detailed kinematic model of NGC~6881
would require unavailable additional observations and, at any rate, the value of 
$\epsilon$ found here is almost entirely constrained by the expansion of the central 
core/ring.  Thus, as long as we restrict ourselves to this specific component, our 
determination provides an accurate estimate of the expansion. From the value of
$\epsilon$ estimated earlier, the angular expansion rate of NGC~6881 can be
calculated as:

\begin{equation}\centering
\dot{\theta} = {{\theta~ \epsilon} \over {\Delta {\rm t}}},
\end{equation}

\noindent
where, $\theta$ is the radius of the ring and $\Delta$t is the time baseline between 
the observations. From the separation of the two peaks seen in the full free-free 
emission images (Figure 1), we estimate that the radius of the ring is 
$0\rlap.{''}73 \pm 0\rlap.{''}01$, and an angular expansion rate of $\dot{\theta} = 2.1 
\pm 0.7~ {\rm mas~yr}^{-1}$ for the central ring of NGC~6881.

\subsection{The distance to NGC~6881}

The distance $d$ to NGC~6881 can be obtained from its angular expansion rate as:

\begin{equation}\centering
\bigg[\frac {d} {\mbox{pc}}\bigg] = 211 \bigg[\frac {v_{\mbox{\tiny{exp}}}} {\mbox{km~s}^{-1}}\bigg] \bigg[ \frac {\dot{\theta}} {\mbox{mas~yr}^{-1}} \bigg]^{-1},
\end{equation}

\noindent where $v_{\rm exp}$ is the expansion velocity of the central ring
along the line of sight. As mentioned previously, the expansion velocity of the 
ring has been estimated from the line splitting of the [NII] by \citet[][]{gm98} to be
$v_{\rm exp}$ = 14.0 km~s$^{-1}$. Other estimates of the expansion 
velocity of NGC~6881 using [OIII] lines resulted in values of 18 km~s$^{-1}$
\citep{rra82} and 16.5 km~s$^{-1}$ \citep{We89}. Thus, we will use an average 
expansion velocity value of  $v_{\rm exp}=$ 16 $\pm$ 2 km~s$^{-1}$. This
leads to a distance $d$ = 1.6 $\pm$ 0.5 kpc; table 2 shows the physical 
parameters of NGC~6881 calculated using this value.  

Distance measurements based on angular expansion can 
potentially be affected by the difference between the pattern
velocity (measured by the expansion) and the true gas velocity
(measured by the radial velocities). Mellema (2004)\nocite{melle04} discusses
this effect in detail, and concludes that it typically results in errors
on the distance of about 20\%. This would add a ``systematic''
error to our measurement of 0.3 kpc. We conclude that the 
distance to NGC 6881 is 1.6 $\pm$ 0.5 $\pm$ 0.3 kpc, where
the first error reflects the uncertainty on the expansion, and
the second error comes from the potential difference between
pattern and material speeds.

The distance found here is in good agreement with that obtained by
\citet{rlgm08},  $d\sim1.5$ kpc. It is also compatible with the
physical association of NGC~6881 with  the nearby HII region Sh 2-109
(see \S 1). When superimposed over the extended  HI emission
(\citealp{cappa96}), NGC~6881 appears projected toward a cavity with a
$v_{\rm LSR}\sim 10$ km~s$^{-1}$,  very close to that of the Cyg OB1
association.  The estimated distance to this association lies in the
range 1.25--1.83 kpc (\citealp{lozin98,schneider06}).The association
Cyg OB 1 is located to the south-west of the Cygnus star-forming
region. Recently, Rygl et al.\ (2010)\nocite{rygl10} measured very accurately the
parallax of the massive star-forming region W75N with the Cygnus
region and found $d$ = 1.32$^{+0.11}_{-0.09}$  kpc. This suggests that
the entire Cygnus complex (which is about 200 pc across) is located to
distances between 1.2 and 1.8 pc. Thus,  our result supports the idea
that NGC~6881 might be associated with the massive star forming region
in Cygnus. Although this may be a simple coincidence,  it may also
 indicate that the progenitor of NGC~6881 could have been a
short-lived intermediate-mass star. Interestingly, some of the
observed characteristics of  NGC~6881 may be readily interpreted in
this scheme. \citet{rlgm08} observed that the southeast H2 lobe of NGC 6881 is less 
extended than its northwest counterpart, and shows a sharp edge. They note that 
a possible explanation for this asymmetry would be an interaction between the 
lobes of NGC 6881 and inhomogeneous interstellar gas. This could occur 
naturally if NGC 6881 were indeed associated with the star-forming region.

\section{Conclusions}

We have presented observations of the planetary nebulae NGC~6881 obtained
using  the VLA at 6 cm (4.8 GHz) at two epochs separated by 9.25
years. Assuming a  self-similar expansion for the ionized gas, we
determine an expansion angular  rate for NGC 6881 of 2.1$\pm$0.7
mas~year$^{-1}$, and a distance of 1.6 $\pm$ 0.5 kpc $\pm$ 0.3 kpc, where
the first error reflects the uncertainty on the expansion, and
the second error comes from the potential difference between
pattern and material speeds. This places
NGC~6881 at the same distance as the HII region Sh 2-109  and the
Cygnus star-forming region.

\section{Acknowledgments}
L.G., Y.G.\ and L.L.\ acknowledge the support of DGAPA, UNAM and
CONACYT (M\'exico).  L.L.\ is indebted to the Guggenheim Foundation
for financial support. D.T. acknowledges support from the Japan
Society for Promotion of Science (project ID: 22-00022). We gratefully
acknowledge useful discussions with L. F. Rodr\'{\i}guez;  this paper
would not have been the same without his comments. We thank the official MNRAS referee (Myfanwy Lloyd) and an
anonymous unofficial referee for their helpful comments on 
our manuscript. This research has
made use of the SIMBAD database, operated at CDS, Strasbourg, France.

\bibliography{biblio}
\bibliographystyle{mn2e}

\label{lastpage}
\end{document}